\documentclass[twocolumn,showpacs,preprintnumbers,amsmath,amssymb]{revtex4}

\usepackage{graphicx}
\usepackage{dcolumn}
\usepackage{bm}

\begin{document}


\title{Landauer Bound for Analog Computing Systems}

\author{M. Cristina Diamantini}
\altaffiliation[Also at ]{Theory Division, CERN, CH-1211 Geneva 23, Switzerland}
\email{cristina.diamantini@pg.infn.it}
\affiliation{
NiPS Laboratory, INFN and Dipartimento di Fisica, University of Perugia, via A. Pascoli, I-06100 Perugia, Italy
}

\author{Luca Gammaitoni}
\email{luca.gammaitoni@nipslab.org}
\affiliation{
NiPS Laboratory and Dipartimento di Fisica, University of Perugia, via A. Pascoli, I-06100 Perugia, Italy
}

\author{Carlo A. Trugenberger}
\email{ca.trugenberger@bluewin.ch}
\affiliation{%
SwissScientific, chemin Diodati 10, CH-1223 Cologny, Switzerland
}


\date{\today}

\begin{abstract}
By establishing a relation between information erasure and continuous phase transitions we generalise the Landauer bound to analog computing systems. The entropy production per degree of freedom during erasure of an analog variable (reset to standard value) is given by the logarithm of the configurational volume measured in units of its minimal quantum. As a consequence every computation has to be carried on with a finite number of bits and infinite precision is forbidden by the fundamental laws of physics, since it would require an infinite amount of energy.

\end{abstract}
\maketitle


In trying  to resolve the puzzle of Maxwell's demon, Szilard was the first to realize that the dynamics of a physical system has implications in terms of information processing.
Landauer \cite{lan}, building on the work developed by Boltzmann, Gibbs and Shannon, argued that the erasure of one bit of information has an entropic load that costs at least $k T \ln2$ energy (where $T$ is the temperature and  $k$ the  Boltzmann constant). This limit  is usually called the \emph{Landauer bound} and the erasure process is addressed as \emph{Landauer reset}, where the initially uncertain logical value of a binary switch is set to a given value (logic zero or one).  
The key realization of Szilard and Landauer is that information can only be processed by computers, i.e. physical systems, and thus it is subject to the laws of thermodynamics: this is the content of the famous statement that Òinformation is physicalÓ. 
This connection between information theory and thermodynamics, advocated by Landauer as a "principle", has been recently experimentally verified in a work by Berut  et al. \cite{cil}, and afterwards also by Roldan et al \cite{Parrondo}.
In \cite{neur} we noted that the reverse is also true: every physical state that contains a certain amount of order can, in principle, be used for computation since it can encode information bits, regardless of its elementary components, and established a connection between the ordering process in continuous phase transition and the Landauer bound.

The Landauer  principle \cite{lan}, in its original formulation, applies only to digital computing systems in which information is represented in discrete units, i.e. number of bits. Here we discuss the extension of the relation between entropy production and information erasure to physical systems implemented as analog computers, where information is a continuous variable. The key idea to generalise the Landauer principle to analog computers is based on the relation between the second principle and entropy production during continuous phase transitions\cite{neur}.

The Landauer principle is based on the connection between the information-theoretic Shannon entropy and the Boltzmann-Gibbs (thermodynamic) entropy, whose reduction is associated to a minimum energy expenditure via the second principle of thermodynamics \cite{lan,seifert,cil,parrondo}. 
 For a digital computer (i.e. a system that can assume a discrete, finite number of logic states $s_i, i = 1... M$) this connection  is based on the identification of a set of $M$ different logic states that are associated with a partition of the configurational space  $\Omega$ of the computing system. When the contribution of the conditional entropy (that represents the contribution of the different microstates inside each $\Omega_i$) is zero \cite{saga,chiuchiu}, the Shannon entropy is equal to the  thermodynamic entropy generated when the reset is performed. In this case,  the variation of the information (Shannon) entropy $\Delta S_S$ between the final state, in which the system is in a specific state with probability 1 and has thus $S_S =0$, and the initial state in which the systems can be in any one of the possible states $s_i$ with probability $p_i$  is given by
\begin{equation}
\Delta S_S =\sum_{i=1}^{M} p_i\ln{p_i} \label{entd}.
\end{equation}
The continuous generalisation of the Shannon entropy is defined as
\begin{equation}
S^{\rm cont}_S = - \int_{x\in M} p(x)\ln p(x) dx\ , \label{entc}
\end{equation}
\cite{shannon,ihara} where $p(x)$ is the probability distribution of the relevant degree(s) of freedom. As in the digital case we have to show that this corresponds to the thermodynamic entropy generated when the erasure is realised in a physical system. In this analog setting, however, there is an additional problem: the information-theoretic continuous Shannon entropy requires an appropriate regularisation that adapts the dimensional character of the relevant degree of freedom to the dimensionless quantity considered in the probability density $p(x)$. This is because the continuous extension of the Shannon entropy, contrary to the discrete entropy,  which is an absolute quantity, is not invariant under change of coordinates \cite{ihara}.
To cure this problem, Jaynes \cite{jay} proposed to modify eq. (\ref{entc}) by introducing  an invariant factor $p_0 (M)$  that represents the density  of the discrete distribution that gives $p(x)$ in the continuum limit: 
\begin{equation}
S^{\rm cont}_S = - \int_{x\in M} p(x)\ln { p(x) \over p_0(M) } dx \  \label{entcm}\ .
\end{equation}
The factor $p_0 (M)$ will arise naturally in the physical realisation of the continuous Landauer reset that we propose and, as we will show, it represents the minimum quantum of configuration volume of the physical system. This will also cure the problem of the classical continuous entropy which can be negative and divergent \cite{wehrl}.

In \cite{neur} it has been shown  that continuous phase transitions, characterized by an order parameter,  for systems exhibiting  a discrete symmetry can also be viewed as information erasure by resetting a certain number of bits to a given value. This connection has been studied for a system  in which stochastic bits are represented by neural networks \cite{mul}.  Such systems typically undergo a phase transition from a disordered phase to an ordered phase with all the bits taking a predefined value when the fictitious temperature governing the stochastic dynamics is lowered. This phase transition is characterized by an order parameter that increases from the value $0$ at the critical temperature to $1$ at zero temperature.  Lowering the fictitious temperature thus is akin to an erasure process that admits possible errors for the $M$ bits encoded in the network.  We will consider the case of an efficient Landauer erasure. In this case the variation of the entropy of the system between the final and the initial states determines the heat dissipated. The heat  produced is the measurable quantity.
We will thus assume  that all work done by lowering the temperature during the transition goes into lowering the entropy of the system, with no additional dissipation. So the entropy difference between the disordered state and the perfectly order states represents the total entropy production during the transition.
It was shown in \cite{neur} that the entropy difference between the two phases is  governed by the same generic information-theoretic expression derived in \cite{gam1,sagawa,gam2} for the generalised Landauer limit in presence of errors. In \cite{neur} the symmetry was a classical $Z_2$ symmetry but the extension to a classical $Z_{(2n + 1)}$ symmetry, with $n = 1/2, 1 , 3/2.....$ (note that the Ising model $Z_2$ corresponds to $n = 1/2$) is straightforward and gives:
\begin{equation}
 - {\Delta S \over k N} =  {\left( S(T_C) - S(0) \right) \over k N} = \ln (2n +1) \ . \label{one}
 \end{equation}

We will thus consider the ferromagnetic Heisenberg model and the breaking of $O(3) \rightarrow O(2)$, that corresponds to considering the reset of a spin (encoding the analog information)  to a standard value. Clearly in this case the spin can take all values on a sphere of unit radius and we have thus a continuum, infinite number, instead of a discrete, finite number, of possible values.

The 3-dimensional ferromagnetic Heisenberg model for $N$ spins is described by the Hamiltonian
\begin{equation}{\cal H} = - {J\over 2} \sum_{<i,j> } s_i \cdot s_j -  H \sum_i s_i  \ ,\label{snc}
\end{equation}
with $J > 0$, $H$ a constant external magnetic field in the $\hat z$ direction  and $|s|^2 =1$. 
We will consider the mean field approximation of this model.
Since this is a very well known problem  \cite{negele}, we will not go trough the derivation but just give the interesting results for our problem.
The model undergoes a phase transition at a critical temperature $T_C$  from a disordered phase (above $T_C$) to an ordered (ferromagnetic) phase (below $T_C$). This phase transition is characterized by an order parameter, the mean magnetization $m$, increasing from the value 0 at the critical temperature to 1 at zero temperature. Lowering the  temperature, thus, is akin to an erasure process where a continuos spin is reset to a standard value due to the average magnetic field generated by all other spins. The magnetization $m$ plays the role of a quantifier for the error probability in the reset operation\cite{gam1}.

The idea of our derivation is to compute the entropy variation during the transition from $T = T_C$ to $T = 0$ and identify it with the Shannon entropy of the erasure process in the analog computing system, where information is encoded in the spin orientation. In fact, we have already shown that for the Ising model, corresponding to the reset of one bit in a digital computing system, this gives exactly the Landauer bound \cite{neur}. For the case of a continuos variable, like an $O(3)$ spin, this will thus give the natural analog generalisation of the Landauer bound to an analog computing system. 
To complete this program we must compute:
\begin{equation}
 - {\Delta S \over k N} =  {\left( S(T_C) - S(0) \right) \over k N} \ , \label{two}
 \end{equation}
 which is the subject of what follows.

Defining the  magnetization $m\equiv (1/N) \sum_i <s_i>$, where $ < .... >$  denotes the thermal average,  we obtain the expression for the mean field Hamiltonian:
\begin{eqnarray}
{\cal H}_{\rm mf} &&= - J \sum_{<i,j> } s_i H_{\rm eff}  + {J N m^2\over 2}  \ , \nonumber\\  
 H_{\rm eff}  &&= (J m + H ) \ , \label{smf}
 \end{eqnarray}
and the partition function (in the limit $H \rightarrow 0$)
\begin{equation}
 Z =  \exp \left[ \beta {J N 3  m^2\over 2} \right]  \int d^3 s \delta (s^2 -1) \exp \left[ \beta J \sum_{i = 1}^N m \cos \theta_i\right]  \ , \label{fdp}
\end{equation}
where $\theta_i$ is the angle between the spin and the $\hat z$ direction and  $\beta = (kT)^{-1}$.
The free energy is thus given by:
\begin{equation}
{F\over N} =  {J m^2 3 \over 2} - {1\over \beta} \ln \left[ {4 \pi \sinh  \beta m J  \over  \beta m J }\right] \ , \label{fed}
\end{equation}
while the entropy is:
\begin{equation}
{S\over k N}=  \ln\left[ {4 \pi \sinh  \beta m J  \over  \beta m J }\right] - \beta m J  L (\beta m J)\ , \label{emd}
\end{equation}
where $L(x) = \coth x - 1/x$ is the Langevin function.

In the limit $m \rightarrow 0 \  (T = T_C)$ we see that
\begin{equation}
{S(T_C)\over k N}  =  \ln (4 \pi) \ . \label{vsf}
\end{equation}
As we anticipated, the $m=0$ value corresponds to the maximal entropy condition that, as expected, is expressed in terms of the logarithm of the volume of the configuration space (the area of a sphere of unitary radius). When we explore the perfect reset condition $m \rightarrow 1  (T = 0)$ however, we find:
\begin{equation}
{S(T\rightarrow 0)\over k N} \rightarrow  - \infty\ . \label{dt}
\end{equation}
A negative divergent entropy is a problem that plagues several classical systems, like the harmonic oscillator, that do not respect the third law of thermodynamics \cite{hng}. 
In order to overcome this problem  we consider the classical Heisenberg model as the proper classical limit of its quantum counterpart. A negative divergent entropy at $T=0$ would imply an infinite amount of work necessary to align the spin (reset).

In the mean field approximation, the quantum  ferromagnetic Heisenberg model is  the model describing a system of quantum, non-interacting spins $s_i$  with $(2s + 1)$ components in a external magnetic field $H_{\rm eff}$. 
This is  again  a very well known problem \cite{pat}, with Hamiltonian and partition function given respectively by:
\begin{equation}
{\cal H} = - H_{\rm eff} \sum_{i=1}^N s_i \ ; \ Z = \left[\sum_{-s}^s \exp \beta H_{\rm eff} s\right]^N \label{meq}\ .
\end{equation}

In the limit of vanishing external magnetic field the entropy reduces to:
\begin{equation}
{S\over k N} = \ln \left[ {\sinh  (1 + {1\over 2s} ) \beta  m J  s \over  \sinh  { \beta m J s \over 2s} }\right]   - \beta m  J s B_s (\beta m J)\ , \label{emd}
\end{equation}
with: 
\begin{equation} B_n (x) =  {2n + 1 \over 2n} \coth \left( {2n + 1 \over 2n} x \right) - { 1 \over 2n} \coth \left( { 1 \over 2n} x \right) \label{ber}
\end{equation} 
the Brillouin function.
In this case, the entropy in the limit $T \rightarrow 0$ has the correct behavior:  ${S (T=0) \over k N}  = 0$.

The classical limit of the quantum Heisenberg model  \cite{leff} is obtained by properly distributing the $(2s+1)$ values of the quantum spin on the classical sphere of area $4\pi$. To this end, following \cite{leff},  we first rescale the spin 
$s \rightarrow s / s_{\rm max}$,   and introduce a density $\Delta$ such that:
\begin{equation} \Delta \left( 2s_{\rm max} +1 \right) \rightarrow 4 \pi \  \ {\rm for} \ \ s_{\rm max}  \rightarrow \infty  \ , \Delta \rightarrow 0 \ \   . \label{con}
\end{equation}
The rescaling of the spin was introduce to ensure the existence of the infinite spin limit (see \cite{leff} and references therein) and $s_{\rm max}$ represents the highest  weight of the representation in the quantum case.
The density $\Delta$ is  the minimum state volume (i.e. the minimum area in the unitary sphere occupied by our spin) allowed by the Heisenberg principle and is thus
\begin{equation} \Delta = {1\over 2 }\hbar^2 s_{\rm max} \ ;   \hbar s_{\rm max}  \rightarrow  s_{\rm clas} \ \ 
{\rm for} \  s_{\rm max} \  \rightarrow \infty  \ , \hbar \rightarrow 0 ,
\label{contl}
\end{equation}
with $s_{\rm class} = 1$ in our case.
Correspondingly, we  thus define the regularised entropy as:
\begin{eqnarray}
{S\over k N} = &\ln{4 \pi \over (2s_{\rm max} + 1)} \left[ {\sinh  (1 + {1\over 2s_{\rm max}} ) \beta   m J   \over  \sinh  { \beta   m J  \over 2s_{\rm max}} }\right] \nonumber \\
 & - \beta  m J  B_{s_{\rm max}} (\beta m J)\ , \label{sreg}
\end{eqnarray}

At the critical temperature, in the $s_{\rm max} \rightarrow \infty$ limit we have again as before
\begin{equation}
{S (T=T_C) \over k N}= \ln (4 \pi) \ . \label{vsr}
\end{equation}
However,  when $T \rightarrow 0$ we now obtain:
\begin{equation}
{S(T\rightarrow 0) \over k N} =  \ln \left({4 \pi \over 2s_{\rm max} + 1}\right)  = \ln {1\over 2 }\hbar^2 s_{\rm max} =  \ln {1\over 2 }\hbar \ , \label{nlz}
\end{equation}
where we used (\ref{con}) and (\ref{contl}). This result tells us that, if we want to avoid the entropy divergence, we cannot actually send $\hbar \rightarrow 0$. 
 In this perspective it is easy to understand that the negative divergence in the classical entropy is due to the fact that the classical probability distribution can be concentrated in regions smaller than this minimum area. $\Delta$ is the factor $p_0(M)$ in eq. (\ref{entcm}). The presence of this factor is not due to the specific model we considered, but is a general consequence of the fact that  the continuous Shannon entropy must be regularised in order to make it invariant upon a change of coordinates.

We are now ready to compute the entropy  variation  during the erasure of a continuous variable represented here by the resetting to a standard value  the continuous spin $s$. 
This is given by:
\begin{eqnarray}
&&-{\Delta S \over k N} = {S(T_C) - S(T=0) \over k N} =  \nonumber \\ && = \ln (4 \pi s_{\rm clas}^2) - \ln {1\over 2 }\hbar^2 s_{\rm max} = \ln {8 \pi s_{\rm clas} \over \hbar} \ \ . \label{rsc}
\end{eqnarray}
which reduces to $\ln {8 \pi  \over \hbar}$ for the Heisenberg model ($s_{\rm class} = 1$).
 This is the analog generalisation of the Landauer bound  where the entropy production during the erasure process is given by the available configuration volume (the area $4\pi$ in this case) measured in units of the minimum quantum of configuration volume $\Delta$.  

There is an interesting consequence that becomes apparent:  even if we start with continuous, analog information, only a finite countably amount of information can be encoded in a physical  system in agreement with \cite{shannon,bek,lloyd}.

 Due to eq. (\ref{entcm}) this is a general results and not only valid for the specific model we considered. The maximum number of possible logic states that can be associated with the system characterized by an angular momentum $L$ is:
\begin{equation}
N_l =  {\rm int} \left({8 \pi L \over \hbar}\right), \label{int}
\end{equation}
where ${\rm int}(a)$ represents the integer part of the number $a$. 
Since any computation that can be performed consists necessarily in the manipulation of the information encoded in the system, this is characterized by a finite number of bits $n = \log_2 (N_l)$.  As an example, if we consider a cube of $5 \times 5 \times 5=125$ atoms \cite{Feynmann} and we assume an angular momentum per atom of the order of $L \approx \hbar$ we have $N_l=3140$ and thus  $n \approx 11.6$ bits, where we assumed that the interactions between the momenta are such that they behave like a single classical momentum.  Under the same assumption, if we consider instead a magnetic nano-dot with $20 nm$ side that contains approximately $200$ million atoms , we have that in such a system we can store up to $n=27.6$ bits, where. If we want to perform with this system a perfect Landauer reset the amount of heat to be dissipated is readily provided by (\ref{rsc}): $Q \geq 19.11 k T$, approximately $30$ times what we would have for a binary system reset.

Based on these considerations, we can conclude that all possible computers have necessarily to be operated with finite precision (i.e. to be digital) and the realization of a truly analog computing system (i.e. with infinite precision) is forbidden by the laws of physics. 

In the remaining part of the paper we want to discuss the possibility to generalize this result to generic symmetry breaking patterns, for example $O(n) \rightarrow O(n-1)$. In this case the computation of the entropy in the mean field approximation is again given by the integral in eq. (\ref{fdp}) with $d^3s \rightarrow d^n s$,
 \begin{equation}
 Z =  \exp \left[ \beta {J N 3  m^2\over 2} \right]  \int d^n s \delta (s^2 -1) \exp \left[ \beta J \sum_{i = 1}^N m \cos \theta_i\right]  \ .\label{fdpn}
\end{equation}
The entropy is thus given by
\begin{equation}
{S \over k N} = \ln \left[ 2^{n\over 2} \pi^{n\over 2}{  I_{{n\over 2}-1} (\beta m J) \over  (\beta m J )^{n\over 2} }\right]  - \beta m J {  I_{n\over 2}(\beta m J) \over  I_{{n\over 2} -1}(\beta m J )}  \ , \label{eon}
\end{equation}
where the  $I_\nu (z) $ are the modified Bessel functions of the first kind.
Again, the entropy production during the erasure of a continuos variable represented here by  the continuous  O(n) spin $s$ is:
\begin{equation}-{\Delta S \over k N} = {S(T_C) - S(0) \over k N} \ .\label{den}
\end{equation}
At $T=T_C$, when $m=0$, eq. (\ref{eon}) becomes:
\begin{equation}
{S(T_C) \over k N} = \ln S_{n-1}  \ ,  S_{n-1} = {2 \pi^{n\over 2} \over \Gamma \left( {n\over 2}\right)}\ ,  \label{eonc}
\end{equation}
$S_{n-1}$ being  the area of the $(n-1)$-sphere of unit radius and $\Gamma (x)$  the Euler Gamma function.
At $T \rightarrow 0$ we have again a negative, logarithmically divergent entropy 
$(S(T\rightarrow 0) / k N) \rightarrow - \infty$.
For the classical  $O(3)$  Heisenberg model in the  mean field  approximation we regularised the entropy by taking the classical limit of the quantum mean field model. In the case of the $O(n)$-symmetric quantum Heisenberg model, however, analytical results are not known \cite{weise}  and, also, the definition of the classical limit is not clear.  
In analogy with the result we have obtained for the $O(3)$ case, and given the corresponding result  eq.(\ref{eonc}), we conjecture that also for the $O(n)$-symmetric case the entropy production during the erasure process will be  given by the available configuration volume, the area of the $n$-sphere, measured in units of the minimum quantum of configuration volume, that in this case will be $\propto \hbar^{n-2}$.
A proof that it is possible to define a positive classical entropy (the Wherl-type entropy) for the Heisenberg model with $SU(n)$ symmetry group,  when restricted to symmetric representation, has been recently proposed by Lieb and Solovey \cite{lieb} using coherent states. Possible applications of this results to the present problem  are under study.

We acknowledge useful discussions with G. Carlotti and S. Lloyd. M.C. Diamantini and L. Gammaitoni acknowledge financial support from
the European Commission (FPVII, Grant agreement no:
318287, LANDAUER and Grant agreement no: 611004, ICT-Energy) and ONRG grant N00014-11-1-0695. MCD acknowledges CERN for kind hospitality during this work.

\end{document}